\documentclass[journal]{IEEEtran}
\usepackage{cite}

\ifCLASSINFOpdf
  \usepackage[pdftex]{graphicx}
\else
  \usepackage[dvips]{graphicx}
\fi
\ifCLASSOPTIONcompsoc
  \usepackage[caption=false,font=normalsize,labelfont=sf,textfont=sf]{subfig}
\else
  \usepackage[caption=false,font=footnotesize]{subfig}
\fi
\usepackage{url}

\begin{document}
%
\title{Elpasolite Planetary Ice and Composition Spectrometer (EPICS): A Low-Resource Combined Gamma-Ray and Neutron Spectrometer for Planetary Science}
%
%
%

\author{K.E. Mesick,
        L.C. Stonehill,
        D.D.S. Coupland,
        D.T. Beckman, S.T. West, S.F. Nowicki, N.A. Dallmann, S.A. Storms, and W.C. Feldman
\thanks{Manuscript received December 13, 2018.} %
\thanks{Research presented in this paper was supported by the Laboratory Directed Research and Development program of Los Alamos National Laboratory under project number 20170438ER.  This research used resources provided by the Los Alamos National Laboratory Institutional Computing Program, which is supported by the U.S. Department of Energy National Nuclear Security Administration under Contract No. 89233218CNA000001.} %
\thanks{K.E. Mesick, L.C. Stonehill, D.D.S. Coupland, D.T. Beckman, S.F. Nowicki, N.A. Dallmann, and S.A. Storms are with Los Alamos National Laboratory, Los Alamos, NM 87545 USA.} %
\thanks{S.T. West is with Arizona State University, Tempe, AZ 85287 USA.} %
\thanks{W.C. Feldman is with the Planetary Science Institute, Tuscon, AZ 85719 USA.} %
\thanks{Corresponding author email address is kmesick@lanl.gov.}}

%
%

\markboth{}
{}


\maketitle

\begin{abstract}
Neutron and gamma-ray spectroscopy (NGRS) is a well established technique for studying the geochemical composition and volatile abundance relevant to planetary structure and evolution of planetary bodies.  Previous NGRS instruments have used separate gamma-ray and neutron spectrometers.  The Elpasolite Planetary Ice and Composition Spectrometer (EPICS) instrument is an innovative and fully integrated NGRS with low resource requirements. EPICS utilizes elpasolite scintillator read out by silicon photomultipliers to combine the gamma-ray and neutron spectrometer into a single instrument, leading to a significant reduction in instrument size, weight, and power.  An overview and motivation for the EPICS instrument, current status of the EPICS development, and a discussion of the expected sensitivity and performance are presented.
\end{abstract}


\IEEEpeerreviewmaketitle

\section{Introduction}

\IEEEPARstart{P}{}lanetary Neutron and gamma-ray spectroscopy (NGRS) from orbiting spacecraft has become a standard technique to measure the geochemical composition of planets or solid bodies such as moons or asteroids. Galactic cosmic rays (GCR) interact with matter in the top tens of centimeters to one meter of planetary surfaces with little to no atmosphere to produce spallation neutrons. Moderation of these neutrons by hydrogen provides a unique signature indicating the presence and abundance of near-surface water that can be in the form of hydrated minerals or water ice. Gamma-rays are produced at characteristic energies either by radioactive decay of natural elements or by collisions or capture of neutrons with elements in the surface material.  These characteristic gamma-rays indicate the presence and abundance of most major and minor rock-forming elements, including H, C, O, Na, Mg, Al, Si, P, S, Cl, K, Ca, Ti, Fe, Th and U.

These measurements are very difficult, requiring good gamma-ray energy resolution, neutron energy sensitivity over ten orders of magnitude, unraveling of the background cosmic radiation, and operation in the space environment under mission resource constraints. Previous planetary missions that have included orbital neutron and/or gamma-ray spectroscopy instruments include Lunar Prospector \cite{Feldman2004} and LRO \cite{Mitrofanov2010} to the Moon, Mars Odyssey \cite{Boynton2004} to Mars, MESSENGER \cite{Goldstein2007} to Mercury, NEAR \cite{Trombka1997} to the asteroid Eros, and Dawn \cite{Prettyman2011} to the asteroids Ceres and Vesta. The previous neutron spectrometers have featured $^{3}$He tubes and/or $^{10}$B-loaded plastic or $^{6}$Li-glass scintillators.  Previous gamma-ray spectrometer instruments have ranged from BGO scintillators with $\sim$10\% full-width half-maximum (FWHM) energy resolution at 662~keV or high-purity Ge (HPGe) with exquisite energy resolution but requiring cryo-cooling systems.  NGRS instruments provide complementary information and in the current generation of instrumentation are separate detectors that may share some components or electronics.

The Elpasolite Planetary Ice and Composition Spectrometer (EPICS) instrument is an innovative combined gamma-ray and neutron spectrometer for planetary science that incorporates elpasolite scintillator material read out by compact silicon photomultipliers to provide significant reduction in size, weight, power, and instrument complexity. Elpasolites are sensitive to and can uniquely distinguish neutrons and gamma rays, enabling a single detector to be used for both gamma-ray and neutron spectroscopy. Elpasolites also provide superior gamma-ray energy resolution over previous scintillator-based gamma-ray spectrometers, enabling improved sensitivity for the detection and quantification of key elements.

\IEEEpubidadjcol
\section{EPICS Instrument}

The EPICS instrument under development at Los Alamos National Laboratory (LANL) is a low-resource, fully integrated NGRS enabled by new scintillator and photodetector technologies. EPICS utilizes elpasolite scintillators which can detect neutrons and gamma-rays within the same detector volume, and silicon photomulipliers which offer significant advantages over traditional readout technology. These technologies combine to offer significantly reduced size, weight, power (SWaP), and high-voltage requirements over previous NGRS instrumentation but with similar or improved performance.

\subsection{Elpasolite Scintillators}

EPICS is based on elapsolite scintillators, which are an emerging new inorganic scintillator material that offers excellent linearity, high light yield, and the ability to detect both gamma-rays and neutrons.  Thermal neutrons are detected through the $^{6}$Li(n,$\alpha$)T reaction, and the chloro-elpasolites offer some fast neutron sensitivity by means of the $^{35}$Cl(n,p)$^{35}$S reaction \cite{Glodo2013,Smith2015}.  Using 95\% enriched $^{6}$Li yields a high efficiency for thermal neutron detection, and densities between 3.3--4.2~g/cm$^3$ provide high stopping power for gamma rays up to 10 MeV.  Cs$_2$LiYCl$_6$:Ce$^{3+}$ (CLYC) is the most mature elpasolite and has been commercially available from Radiation Monitoring Devices, Inc. since 2012. Our team has worked extensively with CLYC and several other elpasolites since early 2010, with particular focus on the application of CLYC for space-based missions.  
\begin{figure}[b!]
\centering
\includegraphics[width=0.48\textwidth]{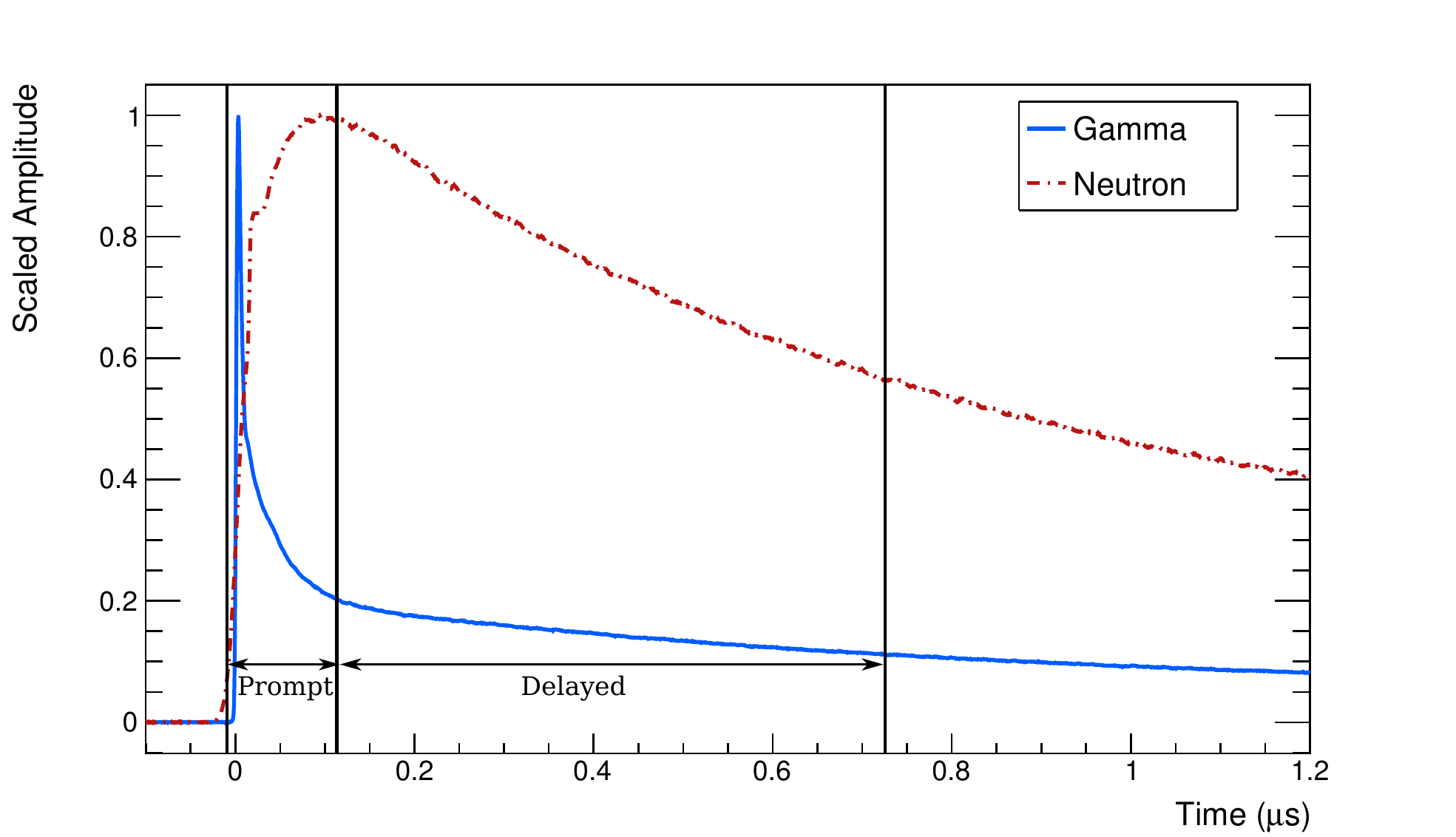}
\includegraphics[width=0.48\textwidth]{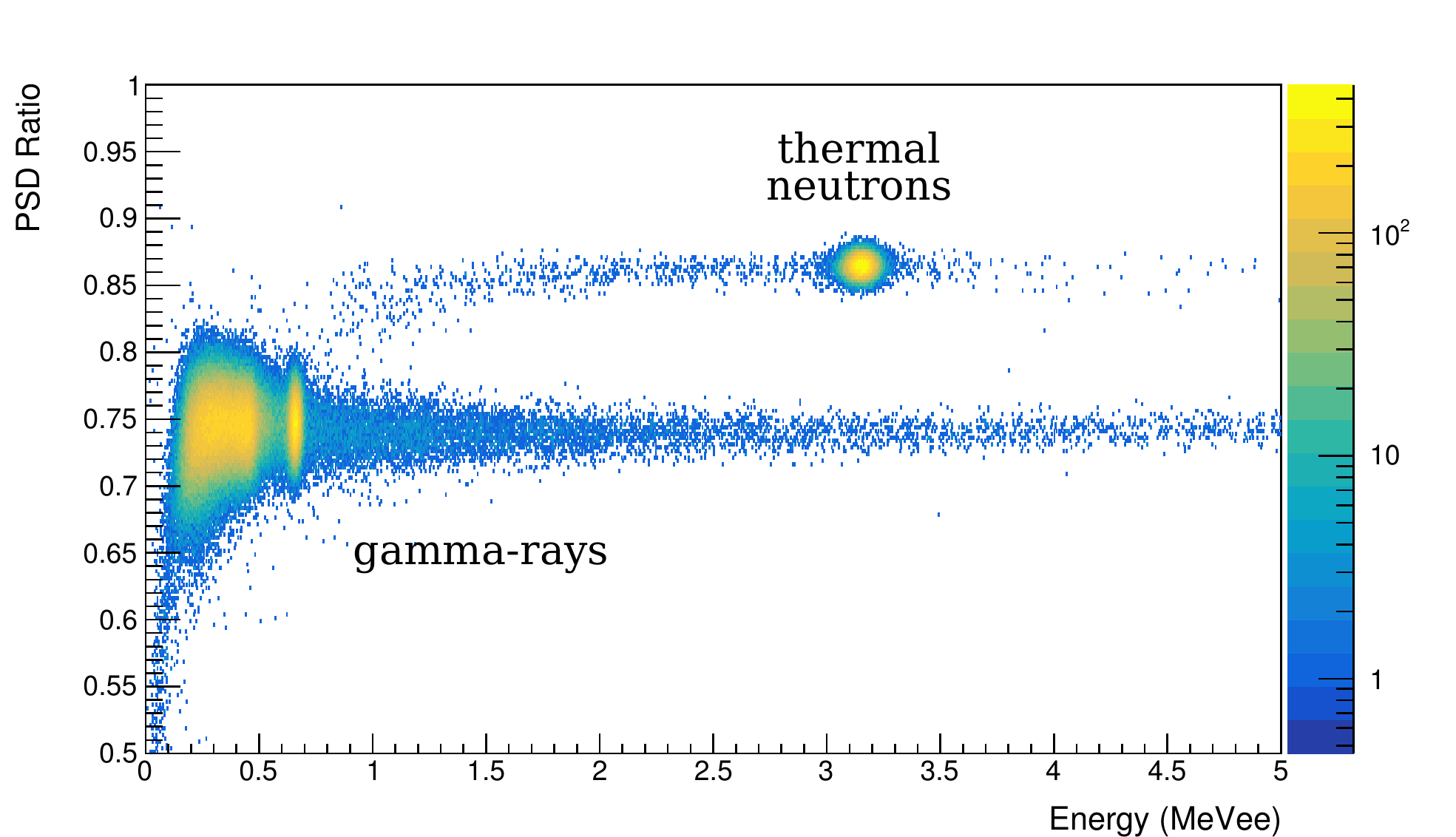}
\caption{(Top) Typical pulse shapes from thermal neutron and gamma-ray radiation in CLYC and (Bottom) Pulse-shape discrimination in CLYC.}
\label{fig:clyc_waves}
\end{figure}

Different scintillation decay times allow gamma rays to be uniquely distinguished from neutrons using pulse-shape discrimination (PSD). Typical pulse shapes from CLYC are shown in the top panel of Fig.~\ref{fig:clyc_waves}.  By forming a ratio of appropriate integrals over different regions of the pulses, gamma rays and neutrons can be robustly separated, as indicated for CLYC in the bottom panel in Fig.~\ref{fig:clyc_waves}.  Thermal neutrons from the $^6$Li capture reaction appear around 3.2 MeVee.  The band to the left of this peak are fast neutrons detected through the $^{35}$Cl(n,p)$^{35}$S reaction.  A PSD figure of merit (FOM) is calculated by taking the difference in the gamma ray and neutron centroid in the PSD ratio divided by the sum of the FWHM of those peaks.  CLYC has the best PSD FOM among the current elpasolites with measurements as good as 4.55 reported \cite{Lee2012} and an energy resolution as good as 3.9\% FWHM at 662~keV \cite{Glodo2011}.  Another promising elpasolite is Cs$_2$LiLaBr$_6$ (CLLB) which offers a higher light yield than CLYC and therefore an energy resolution as good as 2.9\% \cite{Glodo2011}.  

Elpasolites currently do not have any space heritage, however, two space instruments utilizing CLYC scintillator are currently being built for flight: a LANL-developed national security instrument that will operate at geosynchronous orbit \cite{Coupland2016} and the Arizona State University developed LunaH-Map CubeSat mission which is planned for NASA's EM-1 launch to the Moon \cite{Hardgrove2016}.

\subsection{Silicon Photomulipliers}\label{sec:sipms}

\begin{figure}
\centering
\includegraphics[width=0.28\textwidth]{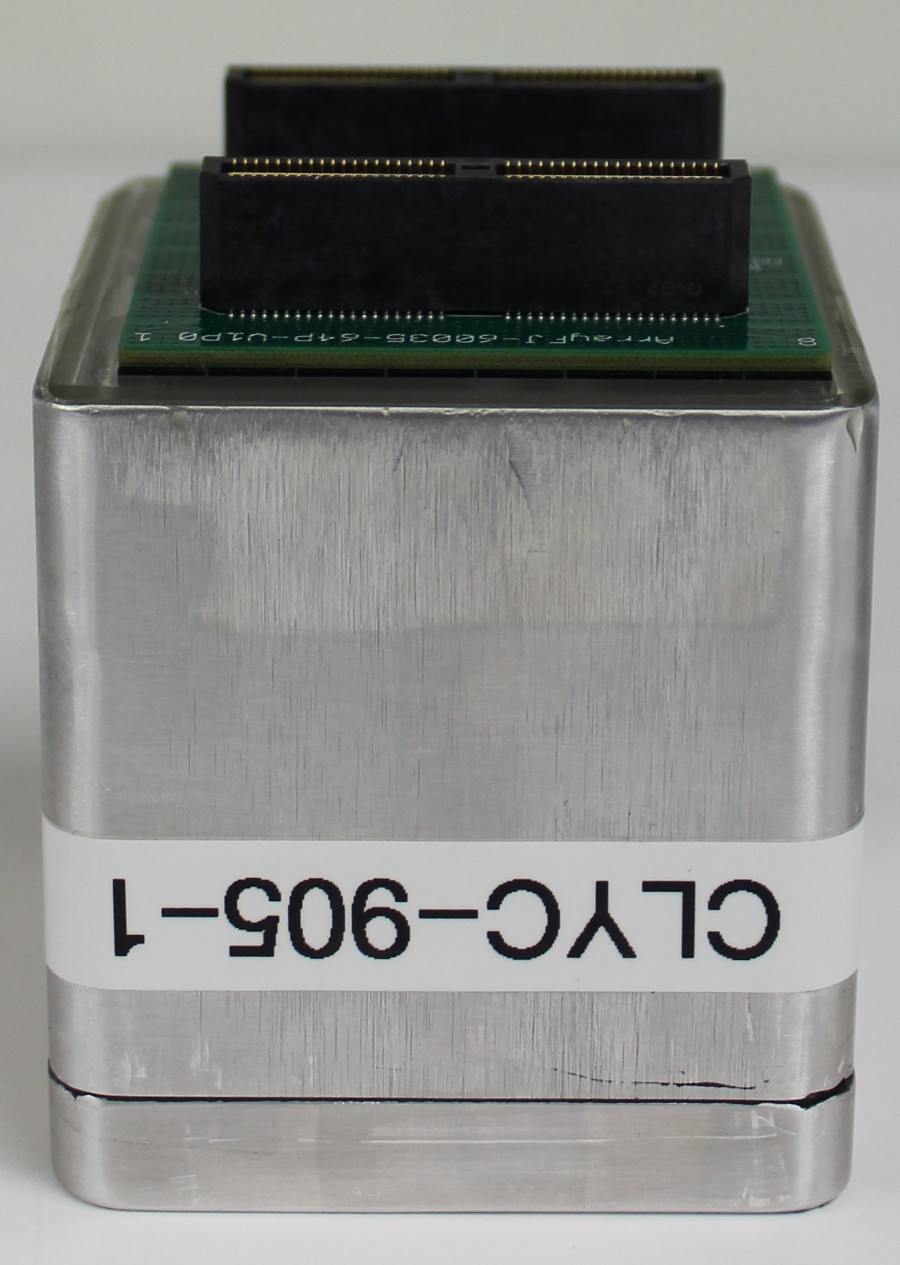}
\caption{Picture of 5 cm $\times$ 5 cm $\times$ 5 cm CLYC crystal with 64-element SiPM array, part of the EPICS prototype.}
\label{fig:clycsipm}
\end{figure}

Scintillator-based radiation detectors require transduction of scintillation light to measurable electrical signals. Silicon photo-multipliers (SiPMs) offer similar gain and noise performance to traditional photo-multiplier tubes (PMTs), but are smaller and lighter weight, and require tens of volts of bias rather than kV-scale bias typical of
PMTs. Newer SiPMs with good response at blue wavelengths have shown excellent performance
with elpasolites \cite{Mesick2015}. 

Arrays of SiPMs are required to read out large crystals; controlling the required channel count while maintaining pulse shape integrity has been a major focus of this effort.  The primary challenge of using a SiPM array to read out large scintillator crystals is minimizing the channel count while preserving the PSD performance.  Simple summing increases capacitance and destroys pulse shape information.  

\begin{figure}
\centering
\includegraphics[width=0.38\textwidth]{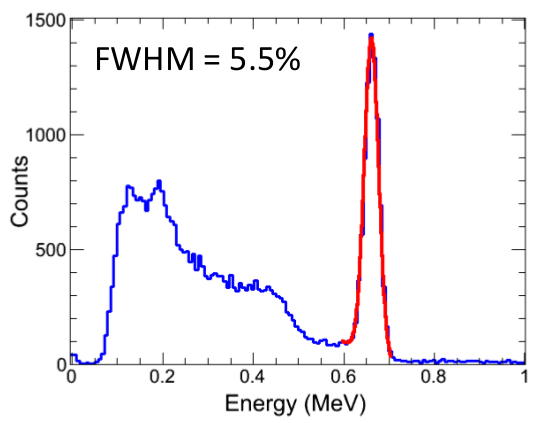}
\includegraphics[width=0.38\textwidth]{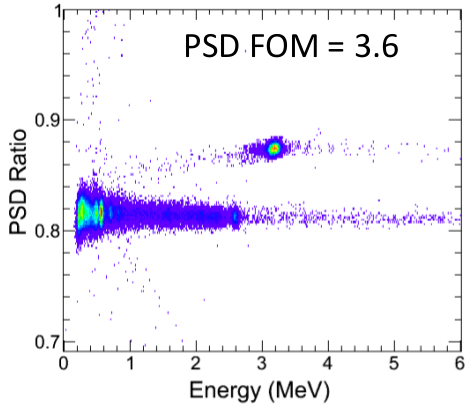}
\caption{(Top) Gamma-ray energy spectrum from a Cs-137 source collected with the CLYC crystal and SiPM array, (Bottom) PSD from the CLYC crystal and SiPM array using Cf-252 and Th-232 sources.}
\label{fig:sipm}
\end{figure}

\begin{figure}
\centering
\includegraphics[width=0.48\textwidth]{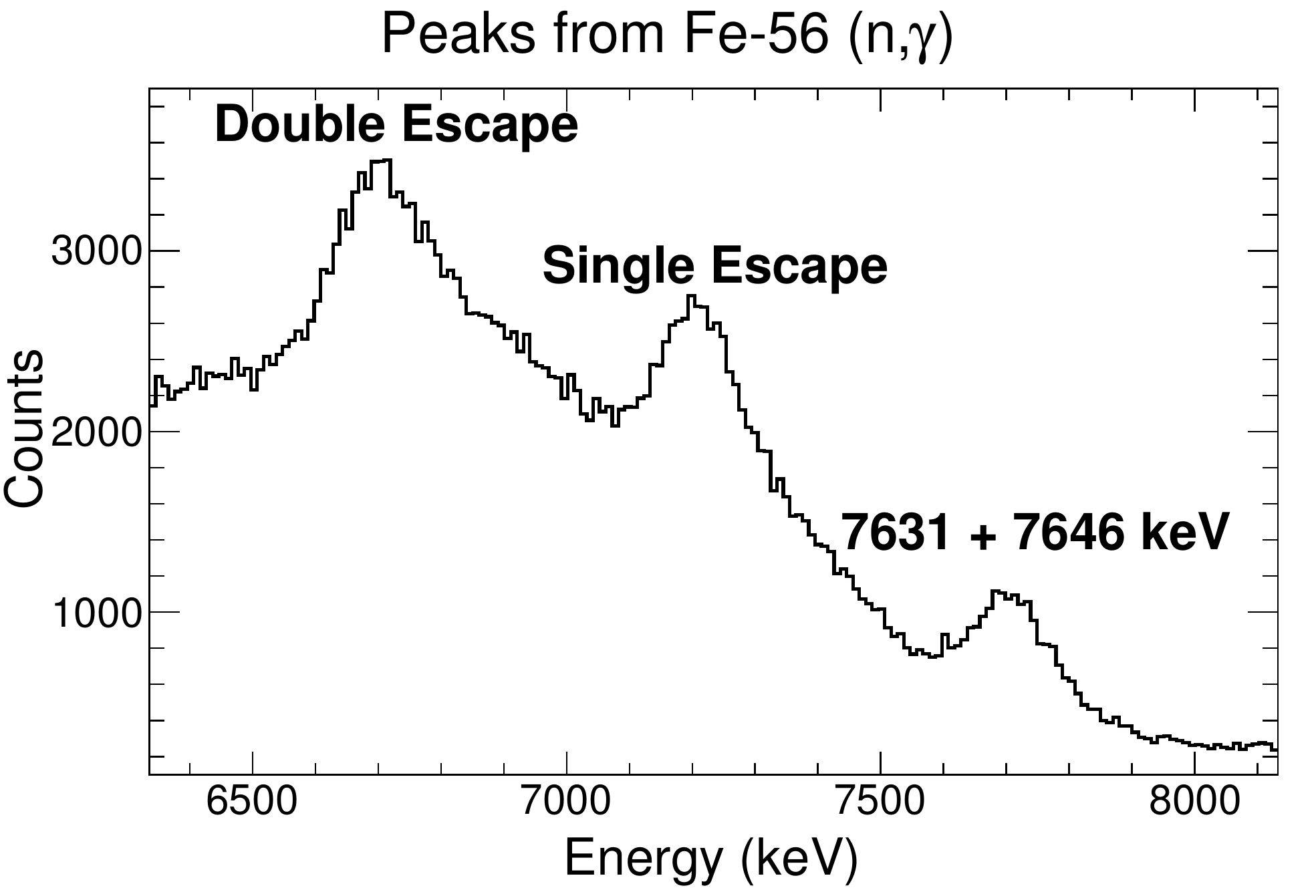}
\caption{Measurement of Fe-56(n,$\gamma$) gamma-ray lines near 7.6 MeV, including the single and double escape peaks, from the 5 cm $\times$ 5 cm $\times$ 5 cm CLYC read out with a SiPM array.}
\label{fig:iron}
\end{figure}

We have optimized an amplification and combing circuit that reads out a 64-element SensL J-series array with 6$\times$6~mm$^2$ elements on a 5 cm $\times$ 5 cm $\times$ 5 cm CLYC crystal (shown in Fig.~\ref{fig:clycsipm}) with only four amplifiers and a single readout channel.  The performance of this system approaches results for energy resolution and PSD FOM when the same crystal is read out with a 2'' super bi-alkali PMT.  Laboratory measurements, shown in Fig.~\ref{fig:sipm}, demonstrate a 5.5\% FWHM energy resolution at 662~keV and a FOM of 3.6, yielding excellent PSD and spectroscopy.  We have also measured the linearity of the CLYC + SiPM array and demonstrated excellent linearity up to 8 MeV and can cleanly detect the 7.6 MeV iron doublet and associated single and double escape peaks from $^{56}$Fe(n,$\gamma$) (see Fig.~\ref{fig:iron}).

\subsection{EPICS Design}\label{sec:design}

\begin{figure}[b!]
\centering
\includegraphics[width=0.48\textwidth]{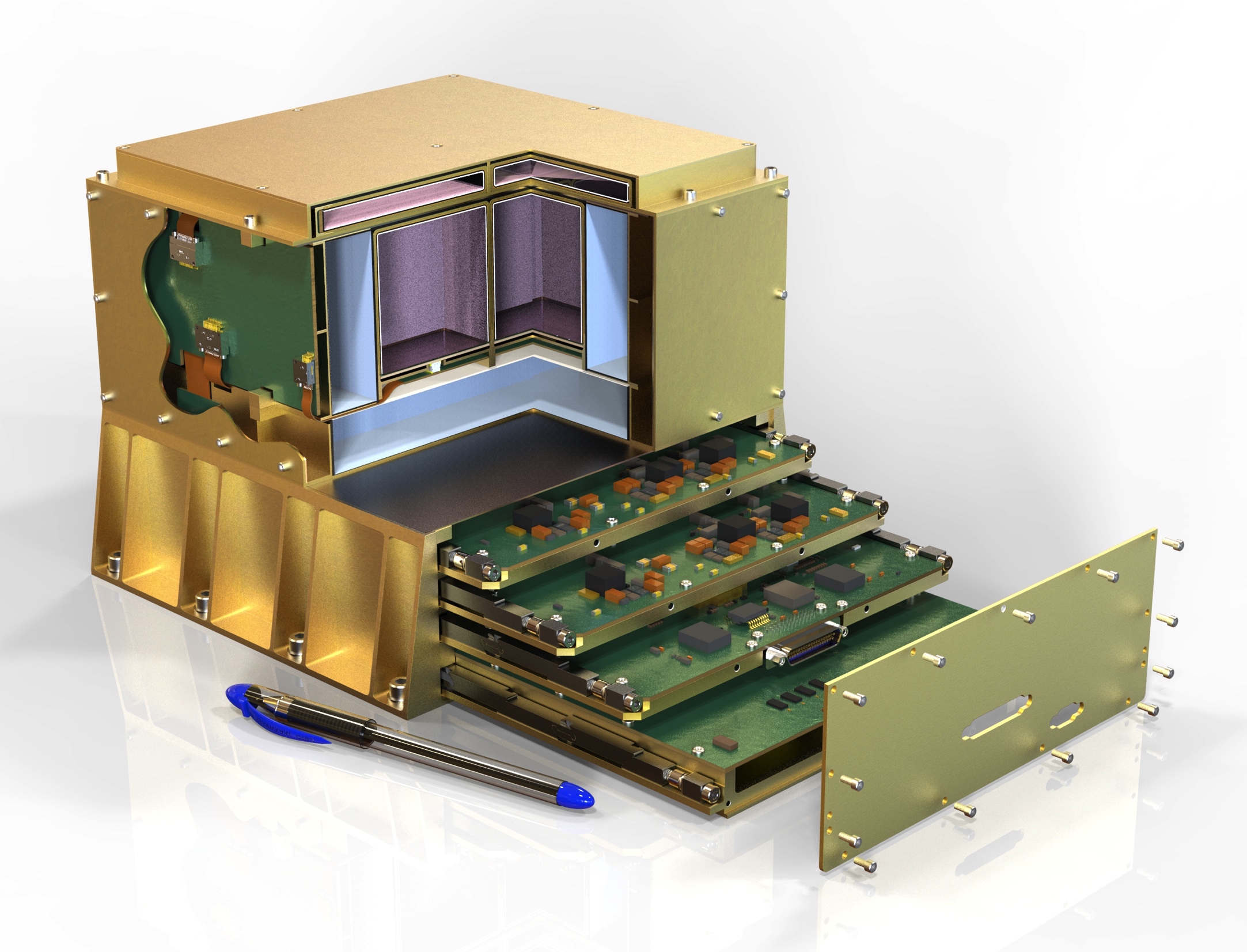}
\caption{Rendering of the EPICS instrument.  The nadir direction is intended to be up in this figure.}
\label{fig:epics_rendering}
\end{figure}

The EPICS instrument baseline utilizes CLYC scintillator, tiled with an array of SiPMs for readout. As the production of other elpasolite materials matures, CLYC can be trivially replaced by other elpasolites if their performance is more desirable. EPICS is designed to be a modular instrument that can be scaled appropriately to meet mission and performance requirements.  Each module is a thin (6~mm) CLYC scintillator for detecting thermal neutrons in front of larger (5 cm cube) CLYC scintillator for detecting epithermal and fast neutrons and gamma-rays. The thin CLYC scintillator is intended to face nadir toward the target body and is backed with 0.63-mm thick Cd to prevent thermal neutrons from the planetary body entering the central volume. The other faces of the larger central CLYC scintillator are surrounded by EJ-200 plastic scintillator, also read out with SiPMs, which acts as a veto for charged-particle backgrounds and is used in coincidence with the central CLYC volume to detect fast neutrons. A rendering of the EPICS instrument with a 2$\times$2 array of these modules is shown in Fig.~\ref{fig:epics_rendering}. With supporting electronics, shielding, and packaging, the instrument in this configuration weighs approximately 7 kg.

EPICS with CLYC is expected to have a gamma-ray energy resolution of 5--6\% FWHM at 662 keV over the entire temperature range of -30$^{\circ}$C to +60$^{\circ}$C typically encountered in space, even after accounting for radiation damage incurred over a five-year interplanetary mission. The performance of an NGRS in the most general sense is expressed as an effective area, which is the efficiency for the instrument to detect the particle of interest multiplied by the detector area. Under the four-module configuration shown in Fig.~\ref{fig:epics_rendering}, the gamma-ray photopeak detection efficiency is $>$7 cm$^2$ at the 7.6 MeV iron doublet lines and $>$25 cm$^2$ at the 609 keV line from uranium decay. The effective area for neutron detection is $\sim$120 cm$^2$ for thermal neutrons, $\sim$30 cm$^2$ for epithermal neutrons, and $\sim$1 cm$^2$ for fast neutrons.  The results shown for the SiPM array readout in Section~\ref{sec:sipms} were collected using the central CLYC volume in a prototype of the instrument, which consists of one of the four modules shown in Fig.~\ref{fig:epics_rendering}.

The processing electronics have been used in previous elpasolite-based instruments built at LANL \cite{Budden2015a,Budden2015b} and are built around the PSD8C ASIC \cite{Engel2009}.  These ASICs support eight channels with three gated integrators (prompt and delayed PSD windows, energy) for each channel, enabling a compact electronics board to easily handle the multi-channel EPICS sensor.

\section{Performance Simulations}

The EPICS instrument is applicable to numerous potential rendezvous and landed missions to airless or near-airless bodies.  The full sized instrument (2$\times$2 array of modules, as shown in Fig.~\ref{fig:epics_rendering}) has been optimized for an orbital rendezvous with a small planetary body, such as an asteroid or moon.  Based on this, we have performed simulations to estimate the instrument requirements and expected sensitivity and performance of EPICS in several potential missions, including an asteroid rendezvous and a mission to the moons of Mars.   While EPICS is being designed in this effort based on these notional missions, smaller version of EPICS (\textit{e.g.} a single module) would also be useful.  For example, EPICS could produce higher-resolution hydrogen maps on the Moon or Mars from a low-orbit spacecraft, balloon, or other flying craft, a topic of interest for human exploration.  EPICS could also access the availability of water and key elements from a small lander on the Moon or Mars relevant to resource utilization.

Simulations of the GCR-induced neutron and gamma-ray albedo were performed using the radiation transport code MCNP6.2 \cite{MCNP}.  The resulting fluxes for a given planetary composition and GCR flux were then used as inputs into a second simulation to estimate the EPICS performance using the radiation transport code Geant4 \cite{geant}.

\subsection{Asteroid Rendezvous}

\begin{figure}
\centering
\includegraphics[width=0.48\textwidth]{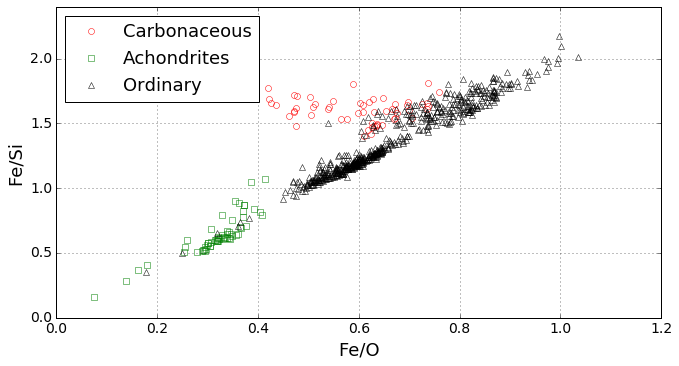}
\includegraphics[width=0.48\textwidth]{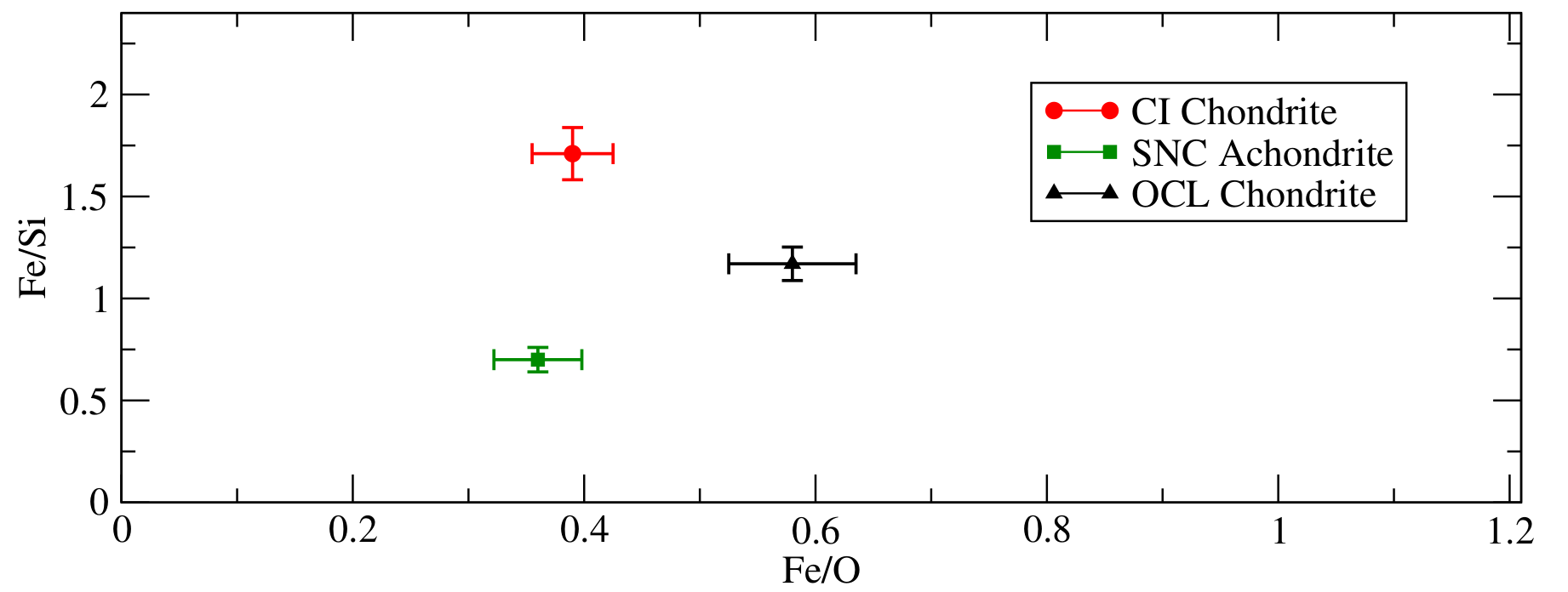}
\caption{(Top) Element ratios Fe/Si versus Fe/O showing discrimination between three groups of asteroid compositions, (Bottom) Simulated uncertainties from a 2-week measurement for these ratios of asteroids within each group.}
\label{fig:asteroid_gamma}
\end{figure}

\begin{figure}
\centering
\includegraphics[width=0.48\textwidth]{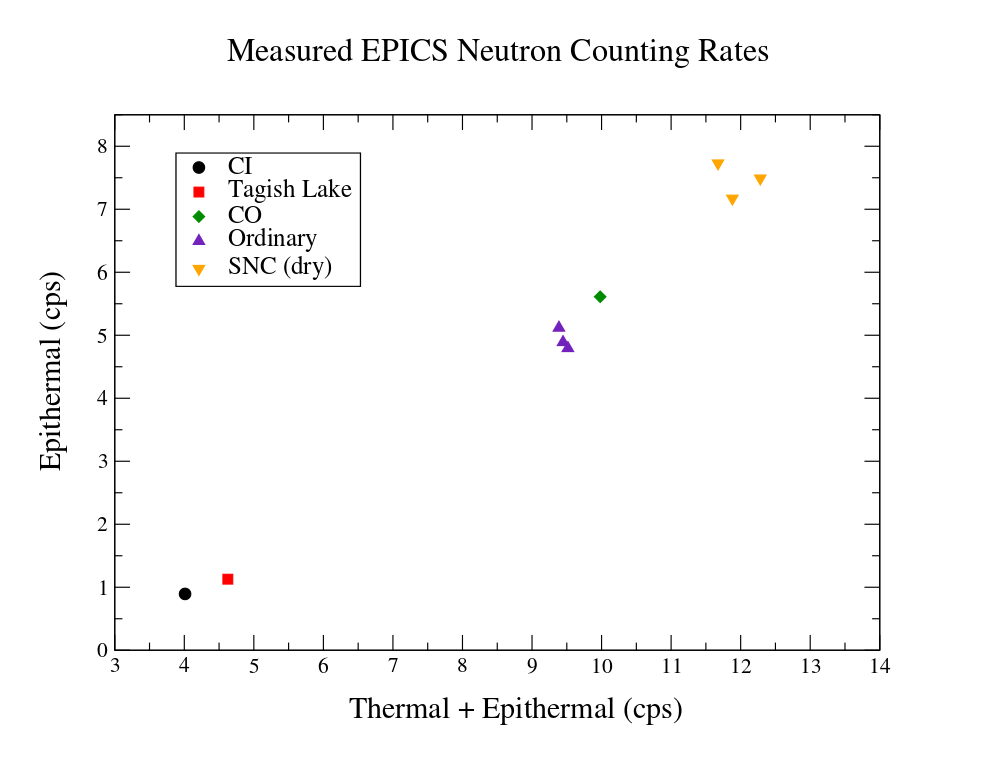}
\caption{Neutron counting rates measured by EPICS for asteroid compositions with H ranging from 0 ppm (SNC's) to 2 wt\% (CI). Statistical uncertainties for 24 hour measurement are smaller than the points.}
\label{fig:asteroid_neutrons}
\end{figure}

An example of the utility of EPICS for geochemical composition measurements is shown in Fig.~\ref{fig:asteroid_gamma}.  In the top panel, elemental ratios for Fe/Si versus Fe/O taken from the Nitler database \cite{Nitler2004} show clear groupings for three different types of asteroid compositions: carbonaceous chrondrites, achondrites, and ordinary chondrites.  Using EPICS to measure Fe (7.6 MeV doublet), Si (1.78 MeV), and O (6.13 MeV), a 2-week measurement at an orbital distance of 1 radius above the surface allows for percent-level statistical uncertainties to be achieved.  This corresponds to the uncertainties shown on the ratios in the bottom panel of Fig.~\ref{fig:asteroid_gamma} for three simulated asteroids within those groups.  The hydration of asteroids can also vary dramatically, and given higher neutron counting rates typically than gamma measurements, precise measurements of the hydrogen abundance can occur more rapidly.  Figure~\ref{fig:asteroid_neutrons} shows simulated EPICS neutron counting rates after a 24 hour measurement, with asteroids containing high amounts of H in the lower left (2\% for CI and 1.5\% for Tagish Lake), to lower H contents (1100 ppm for the ordinary chondrites and 700 ppm for CO), to asteroids with no H (SNCs) in the upper right.  Statistical uncertainties are shown in this plot, and are smaller than the points.  EPICS can easily measure and distinguish ppm-level H enrichments.

\subsection{Martian Moons}

\begin{figure}
\centering
\includegraphics[width=0.48\textwidth]{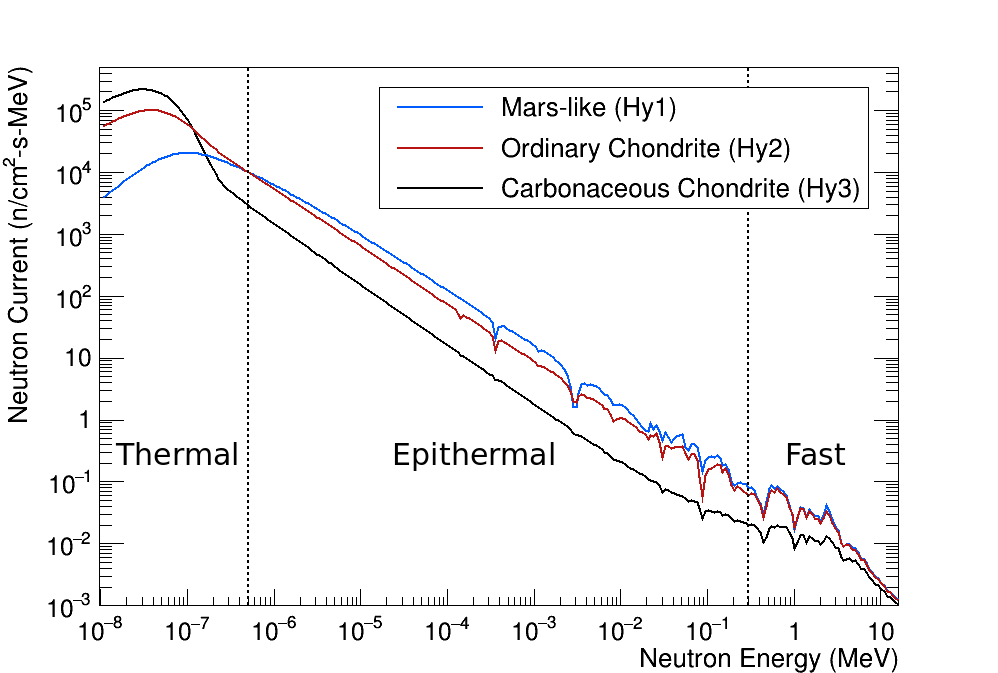}
\caption{The neutron albedo from the martian moons assuming compositions based on three origin hypotheses.}
\label{fig:mmx_neutrons}
\end{figure}

\begin{figure}
\centering
\includegraphics[width=0.48\textwidth]{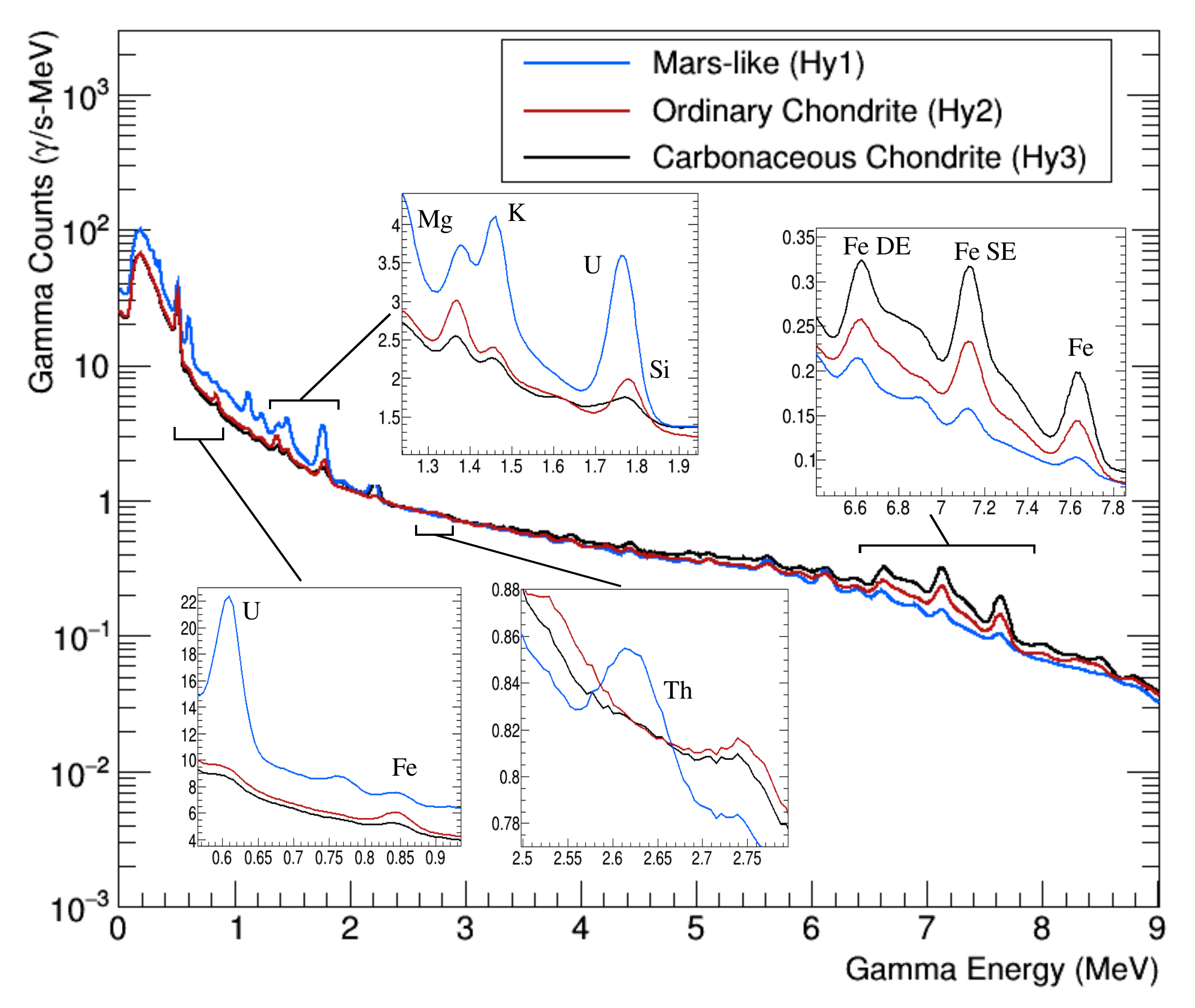}
\caption{Simulated as-measured gamma ray from EPICS assuming three compositions for the martian moons.}
\label{fig:mmx_gammas}
\end{figure}

There are three main hypothesis for how the martian moons Phobos and Deimos were formed.  Hy1) in-situ accretion from giant impact ejecta; Hy2) co-accretion from the proto-martian disk, which would result in ; and Hy3) capture of a primitive asteroid or extinct comet.  Elemental compositions representative of these hypotheses were simulated and Fig.~\ref{fig:mmx_neutrons} shows the neutron albedo at an orbital distance of 2 radii above the moon's surface.  With the effective areas for neutron detection of the EPICS instrument as given in Section~\ref{sec:design}, the three origin hypotheses can be distinguished with significance using the neutron signals in much less than one Earth day in orbit or after several fly-by's.  Key elements that can also be used to distinguish the three origin hypotheses are indicated in Fig.~\ref{fig:mmx_gammas}, which shows the expected gamma-ray spectrum as measured by EPICS, with a 6\% FWHM energy resolution.

\section{Summary}

The EPICS instrument is an innovative neutron and gamma-ray spectrometer for planetary science that fully and optimally integrates the two spectrometers into a single instrument.  This is achieved using elpasolite scintillator for combined neutron and gamma-ray detection in a single volume and compact silicon photomultipliers for readout. 
The low-resource nature of the EPICS instrument makes it beneficial to many future planetary science missions from rendezvous missions to previously unexplored planetary bodies to low-altitude resource mapping missions.

We are in the process of building a prototype of the EPICS instrument and have demonstrated the ability to read out a large CLYC volume with an array of SiPMs and preserve excellent energy resolution and pulse-shape discrimination.  Extensive simulations have been performed of the EPICS instrument response to several potential missions, which show the utility of EPICS to obtain volatile and element abundance measurements, providing critical information for understanding the role of primitives bodies as building blocks for planets and life.


\ifCLASSOPTIONcaptionsoff
  \newpage
\fi

\end{document}